\documentclass[conference]{IEEEtran}
\IEEEoverridecommandlockouts

\usepackage{cite}
\usepackage{amsmath,amssymb,amsfonts}
\usepackage{algorithmic}
\usepackage{graphicx}
\usepackage{textcomp}
\usepackage{xcolor}
\usepackage{subcaption}                  
\usepackage{caption}                     
\usepackage{hyperref}  
\usepackage{mathrsfs}
\newcommand{\etal}{\textit{et. al.}}

\def\BibTeX{{\rm B\kern-.05em{\sc i\kern-.025em b}\kern-.08em
    T\kern-.1667em\lower.7ex\hbox{E}\kern-.125emX}}
\begin{document}

\title{Classification of White Blood Cell Leukemia with Low Number of Interpretable and Explainable Features\\
}

\author{\IEEEauthorblockN{ William Franz Lamberti}
\IEEEauthorblockA{\textit{Center for Public Health Genomics} \\
\textit{Department of Biomedical Engineering}\\
\textit{University of Virginia}\\
Charlottesville, VA, United States \\
william.f.lamberti@virginia.edu}
}

\maketitle
\IEEEpubidadjcol

\begin{abstract}
White Blood Cell (WBC) Leukaemia is detected through image-based classification.  Convolutional Neural Networks are used to learn the features needed to classify images of cells a malignant or healthy.  However, this type of model requires learning a large number of parameters and is difficult to interpret and explain.  Explainable AI (XAI) attempts to alleviate this issue by providing insights to how models make decisions.  Therefore, we present an XAI model which uses only 24 explainable and interpretable features and is highly competitive to other approaches by outperforming them by about 4.38\%.  Further, our approach provides insight into which variables are the most important for the classification of the cells.  This insight provides evidence that when labs treat the WBCs differently, the importance of various metrics changes substantially.  Understanding the important features for classification is vital in medical imaging diagnosis and, by extension, understanding the AI models built in scientific pursuits.  
\end{abstract}

\begin{IEEEkeywords}
White blood cells, explainable artificial intelligence, random forest, image classification
\end{IEEEkeywords}

\section{Introduction}
Acute Lumpocytic Leukemia (ALL) is a disease that affects WBCs.  It is particularly harmful for children and the elderly.  Further, if ALL remains unchecked, it can spread very quickly throughout a patient's body by traveling through their bloodstream \cite{sahlol_efficient_2020}. Thus, being able to accurately classify healthy white blood cells (WBCs) against those with ALL is crucially important.  Further, understanding how AI systems make decisions is crucial for understanding the biological differences between cancerous and healthy WBCs. 

A popular method for the analysis and classification of images is using a convolutional neural network (CNN) \cite{gu_recent_2018}.  This modeling approach is also popular in medical imaging analysis \cite{sahlol_efficient_2020}.  One downside of using a CNN-based approach is that it is difficult to explain and interpret the model's parameters since the CNN models can have millions of parameters.  While some are attempting to make CNNs more explainable and interpretable\cite{samek_explainable_2017}, these approaches still provide little insight to the underlying function estimated and the ability to assign physical meaning to the millions of individual parameters\cite{barredo_arrieta_explainable_2020}.  The explainability and interpretability of a model is important as communities require models used for patients to be interpretable and explainable.  For instance, the European Union (EU) passed the General Data Protection Regulation (GDPR) in 2016 which took effect in 2018 \cite{noauthor_regulation_2016}.  Further, using models with interpretable and explainable features enable analysts to better understand the phenomena of interest.  Thus, using features that are interpretable and explainable enable a deeper understanding of how features interact with one another.  Understanding how these features are used by the model is a key aspect for many questions in science. 

Our proposed approach aims to classify WBCs as healthy or ALL, achieve a high classification performance, provide insight to the important features, and use a relatively low number of parameters.  An overview of our approach is provided in Figure \ref{fig: overview}.  We first processed our images in 1).  Since we used two separate data sources that were collected using different processes, we applied a unified custom segmentation algorithm to treat both datasets equitably.  We then extracted the useful features for shape, color, and texture using the processed images in 2).  Lastly, a random forest (RF) was built using 5-folds cross validation (CV) and evaluated using the validation data in 3).  Our RF model uses interpretable and explainable shape, color, and texture features.  Further, we can describe the importance of each of the variables in the model.  This provides insight into what is useful for determining the type of cell.  We extend the individual variable importance by summarizing the most important group of features.  This process was repeated for both data sets and the models were compared.  Furthermore, our approach outperforms leading approaches by about 4.38\%.  Our GitHub link provides the code to our analysis under separate branches for each respective dataset:\url{https://github.com/billyl320/wbc_luke}

\begin{figure*}[h]
\centering
\includegraphics[width=18cm]{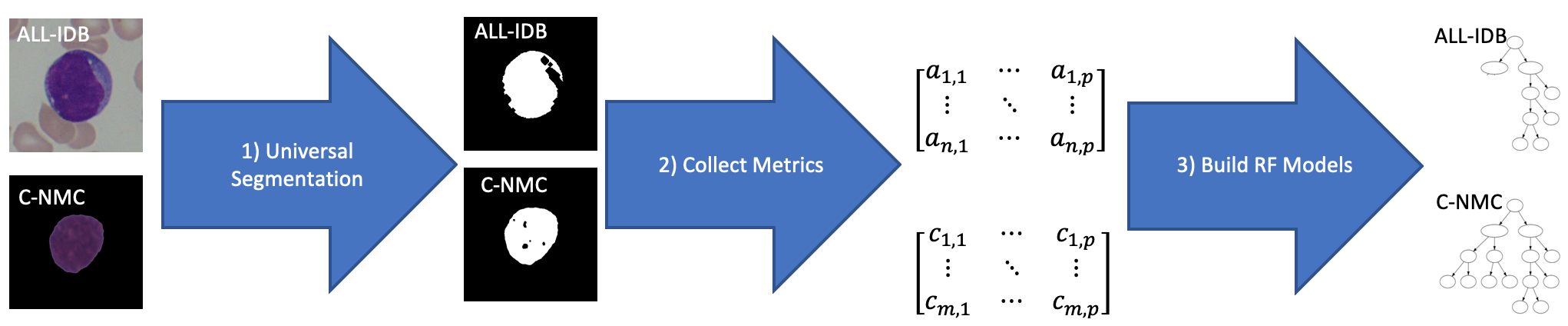}
\caption{ Figure provides overview of our workflow.  The labels ``ALL-IDB'' and ``C-NMC'' were added to the images for visualization purposes.  }
\label{fig: overview}
\end{figure*}

\section{Methods}

\subsection{Dataset description}

The two sources of data used in this manuscript were the publically available and well regarded ALL-IDB\cite{labati_all-idb_2011} and C-NMC\cite{gupta_stain_2017, duggal_overlapping_2016, duggal_sd_layer_2017} datasets.  The ALL-IDB data was provided by the Department of Information Technology - Università degli Studi di Milano.  The image data was captured with a microscope with a Cannon PowerShort G5 camera and are retained in JPG format with 24 bit color depth \cite{labati_all-idb_2011}.  However, the format we received the files were in TIF.  The ALL-IDB source has two datasets, but we will constrain ourselves to the ALL-IBD2 data.  The ALL-IDB2 data contains cropped areas of interest of WBCs that are malignant and healthy.  These cells retain the background and other potential nearby cells.  There are a total of 130 of malignant and healthy cells, for a total of 260 images.  An example from the ALL-IDB2 dataset is provided before the first step in Figure \ref{fig: overview}.

The C-NMC dataset was also used to help broaden the scope of our approach.  This dataset comes from the ISBI 2019 challenge.  The images are saved in the BMP format.  There are a total of 7,272 malignant cells and 3,389 healthy cells for a total of 10,661 images.  The cells were extracted by an expert oncologist.  An example from the C-NMC dataset is provided before the first step in Figure \ref{fig: overview}.  

An important distinction between the ALL-IDB2 and C-NMC datasets is that they both have differing backgrounds.  This is crucially important as the segmenting algorithm we developed can be applied to both datasets without any changes.  This allows for our method to be applied to a much larger variety of data.  Examples of the resulting segmentation results are provided in Figure \ref{fig: overview} after the first step.   

\subsection{Image Preprocessing and Segmentation}

We need to extract the object of interest to make appropriate inferences.  This requires us to perform some image preprocessing.  Since we are using two different datasets that were collected differently, creating a universal segmentation algorithm is non-trivial. We will provide a brief overview of the image operators here, but the exact image operators performed on the image data is provided in the Appendix at the following link: \url{https://github.com/billyl320/wbc_luke/blob/master/appendix.pdf}.  

We first converted all the black pixels to yellow.  We then converted the image to CMYK space.  Next, we convolved the image with a local maximum.  These steps extracted the object of interest from the background.  We then created a contrast stretched image and an equalized histogram image.  We then combined these two images and extracted the object closest to the center of the image.  The resulting image was critically important to extract the features for the analysis.  This algorithm is noteworthy as it is able to properly segment two sets of images that were treated very differently during the data collection stage.  

\subsection{Interpretable Features}

\indent \indent The code to collect the metrics is provided at our GitHub link: \url{https://github.com/billyl320/wbc\_luke}.  Each dataset has is own respective branch and associated code and output.  We selected the 8 shape metrics that had past evidence for outperforming CNNs for other problems related to classifying satellite imagery \cite{lamberti_classification_2020}.  We then collected 14 color metrics from the RGB and CMYK color channels.  We used these color channels since these channels were useful to isolate the cells as performed by Ghane et al. \cite{ghane_segmentation_2017}.  From our extracted object, we also measured the mean and SD of the co-occurance matrix, $P$.  These 2 features extract relevant information regarding texture.  We used the co-occurance matrix as it is a well known approach for collecting texture information \cite{kinser_image_2018}.  The metrics we used are summarized in Table \ref{tab:metrics}.  The exact manner in which we collected these metrics and extended details on these metrics are provided in the Appendix.  However, a shorter summary and description of each of the features is provided in the following sections. 

\begin{table}[ht!]
\centering
\caption{ Table provides the metrics used in this analysis on a given image, $i$.  The first column is the $q^{\text{th}}$ metric, where $q \in \{1, 2, ..., 24\}$. }
\begin{tabular}{c|cc}
  \hline
\begin{math}
  \vec{m}_{q,i}
\end{math} & Metric                   & Feature Type \\ 
  \hline
  1      & White EI                   & Shape  \\ 
  2      & Black EI                   & Shape\\ 
  3      & SP value                   & Shape\\ \\
  4      & $1^\text{st}$ Eigenvalue   & Shape \\
  5      & $2^\text{nd}$ Eigenvalue   & Shape \\
  6      & Eccentricity               & Shape\\ \\  
  7      & Circularity                & Shape \\
  8      & Number of Corners          & Shape\\ \hline 
  9      &Mean R                      & Color \\ 
 10      &SD R                        & Color \\ 
 11      &Mean G                      & Color \\ 
 12      &SD G                        & Color \\ 
 13      &Mean B                      & Color \\ 
 14      &SD B                        & Color \\ \\
 15      &Mean C                      & Color \\ 
 16      &SD C                        & Color \\ 
 17      &Mean M                      & Color \\ 
 18      &SD M                        & Color \\ 
 19      &Mean Y                      & Color \\ 
 20      &SD Y                        & Color \\ 
 21      &Mean K                      & Color \\ 
 22      &SD K                        & Color \\  \hline 
 23      &Mean P                      & Texture \\ 
 24      &SD P                        & Texture \\ 
\hline
\end{tabular}
\label{tab:metrics}
\end{table}

\subsubsection{Shape Features}
The first metrics were the SPs and EIs \cite{lamberti_algorithms_2020}.  The EI is the black and white pixel counts of the cell after the cell is placed in the minimum encompassing circle and then the minimum encompassing square \cite{lamberti_algorithms_2020}.  In other words, this is the area of the cell and the relevant area surrounding the cell.  The SP value is the proportion of the area of the shape relative to the sum of the EI.  The eigenvalues measure the major and minor axes of the shape.  Eccentricity is the ratio of the major axis over the minor one.  These are calculated using the $1^{\text{st}}$ and $2^{\text{nd}}$ eigenvalues of the cells, respectively \cite{kinser_image_2018}.  Circularity measures how circular a given cell is \cite{kinser_image_2018, rosenfeld_compact_1974, proffitt_measurement_1982}.  The number of corners calculates the number of corners in the extracted cell\cite{kinser_image_2018}.  This results in a total of 8 total shape metrics used by our RF model.  
\subsubsection{Color Features}

The metrics for color were the mean and standard deviation of 7 different channels to describe color.  The mean describes the average value of the cell, while the standard deviation (SD) describes how much the color varies within the cell.  Large values for the mean correspond to large amounts of that given color being present in the cell. Small mean values correspond to small amounts of a color's presence.  Large values for the SD indicate that the given color varies throughout the object, while small values indicate that the color is fairly consistent within the object.  The color channels analyzed came from the RGB and CMYK spaces.  The RGB color space is the represents that describe an object's color in terms of red (R), green (G), and blue (B).  The CMYK color space describes the object's color in terms of cyan (C), magenta (M), yellow (Y), and black (K).  The CMYK space is particularly useful for segmenting the WBC from the background \cite{ghane_segmentation_2017}.   This corresponds to a total of 14 color features.

\subsubsection{Texture Features}

The metrics for texture were the mean and SD of the co-occurance matrix in a grayscale representation of the cell.  The co-occurance matrix, $P$, describes how often certain values are next to one another \cite{kinser_image_2018}.  A large mean indicates that grayscale values are nearby a large number of different grayscale values, while a small mean would indicate the opposite.  A large SD indicates that grayscale intensities are nearby many different intensities.  A small SD indicates that grayscale intensities tend to be nearby only a few other intensities.  In other words, smoother images have a small SD while rougher images have a large SD.

\subsection{Modeling}

We will use a random forest (RF) model to classify the healthy and ALL WBCs.  The RF was built using the well established {\tt{caret}} and {\tt{randomForest}} packages in {\tt{R}} \cite{kuhn_caret_2020, cutler_randomforest_2018}.  

Since various communities define validation, training, and testing data differently, we would like to define these terms to be clear.  The {\bf{training data}} is the data used to build the model.  The {\bf{testing data}} is used to evaluate the model built.  The {\bf{validation data}} is the data never used to build the model.  In a random split, we will have the testing and validation data.  In cross validation (CV), we also will have the testing and validation data.  However, the testing data is split into $k$-folds, where each fold takes turns being the testing data while the remaining $k-1$ folds are the training data.  

We first used stratified random sampling to preserve the proportions from the original data when we split the data using a random split.  The strata are the malignant and healthy WBC images.  The random split between the testing and validation data was 80\% and 20\%, respectively.  We chose these proportions for the splits since we wanted to keep our experimental setup similar to experiments done by others.  We then used 5-fold CV on the larger split to determine the number of variables to check when building the RF and estimating of the overall accuracy. For parameter tuning, we used a grid search ranging from 1 to 10 for the number of variables to consider for each tree.  After the number of variables to check was tuned, we used the entirety of the larger split to build a final model.  We then used the smaller 20\% split, the validation data, to confirm that our model produces reasonable results.  

\section{Results}

\subsection{Accuracy and CI for Our RF Models}

Table \ref{tab: acc_ci} provides the 95\% confidence intervals (CIs) for the models on the training and validation data.  Our models are able to provide highly accurate and consistent results on the ALL-IDB2 data.  While our RF model achieved a high level of overall accuracy, our accuracy did drop from the C-NMC training to validation data.  Further, the validation accuracy was not encompassed in the training 95\% CI, suggesting that our RF model slightly overfit to the training data.  

\begin{table}[ht]
\centering
\begin{tabular}{|c|c|c|c|}
\hline
Source & Accuracy & 95\%CI & Data\\
\hline
 ALL-IDB2  & 1.00   & (0.9824, 1.0000) & Training  \\ \hline
 C-NMC     & 1.00   & (0.9996, 1.0000) & Training \\ \hline 
 \hline
 ALL-IDB2  & 1.00   & (0.9315, 1.000)  & Validation \\ \hline
 C-NMC     & 0.8645 & (0.8493, 0.8788) & Validation  \\ \hline
\end{tabular}
\caption{The accuracy and 95\% confidence interval (CI) for the various datasets from our RF model.}
\label{tab: acc_ci}
\end{table}

\subsection{Performance comparison with other approaches}

Tables \ref{tab: all_results} and \ref{tab: c_results} provide the results for our approach compared to other state of the art approaches for the ALL-IDB2 and C-NMC datasets.  Both of these models used the same preprocessing steps to ensure a fair treatment of the data.   Our model outperformed the other state of the art approaches by about 4.70\% and 3.81\% on average for the ALL-IDB2 and C-NMC datasets, respectively. 

\begin{table}[ht]
\centering
\begin{tabular}{|c|c|c|c|}
\hline
Model & Number of Parameters & Model &Accuracy \\
\hline
 \cite{singhal_local_2014}                & 256       & SVM   & 89.72\%   \\ \hline
 \cite{singhal_texture_2016}              & 4096      & Knn   & 93.84\%   \\ \hline
 \cite{bhattacharjee_robust_2015}   & \textbf{8} & Knn  & 95.24\%   \\ \hline
 \cite{singhal_local_2014}                 & 45        & Knn   & 95.67\%   \\ \hline
 \cite{sahlol_efficient_2020}              & 1087      & CNN \& SVM    & 96.11\%   \\ \hline
 \cite{rawat_classification_2017}              & 142    & Ensemble   & 99.2\%   \\ \hline
 \cite{shafique_acute_2018}              & 6.1$\times10^7$  &  CNN      & 99.5\%   \\ \hline
 \textbf{Ours}                            & 24       & RF     & \textbf{100.00\%}   \\ \hline
\end{tabular}
\caption{Comparison with related works for the ALL-IDB2 data.  The best approaches for the smallest number of parameters and highest accuracy are boldened.}
\label{tab: all_results}
\end{table}

\begin{table}[ht]
\centering
\begin{tabular}{|c|c|c|c|}
\hline
Model & Number of Parameters & Model &F1 \\
\hline
 \cite{marzahl_classification_2019}   & $1.1\times10^7$    & CNN        & 86.9\%   \\ \hline
 \cite{ding_deep_2019}                       &  $8.7\times10^7$ & CNN  & 86.7\%   \\ \hline
 \cite{kulhalli_toward_2019}             & $2.5\times 10^7$ & CNN           & 85.7\%   \\ \hline
 \cite{sahlol_efficient_2020}              & 1,115    & CNN      & 87.9\%   \\ \hline
 \textbf{Ours}                            & \textbf{24}    & RF        & \textbf{90.1\%}   \\ \hline
\end{tabular}
\caption{Comparison with related works for the C-NMC data.  The best approaches for the smallest number of parameters and highest F1 score are boldened.}
\label{tab: c_results}
\end{table}

Our approach was able to outperform the other advanced methods by about 4.38\% on average.  Furthermore, our approach used the smallest number of parameters when compared to the other approaches except for one.  This is particularly impressive since the methods to extract the features were the same on both datasets.  Thus, our segmentation and collected metrics used in our approach is applicable to a large variety of images that are collected differently. 

\subsection{Number of parameters comparison}

Our approach used less features than the other approaches.  It was the second smallest for the ALL-IDB2 data and the smallest on for the C-MNC data.  For both datasets, the best prior solutions used a large number of parameters.  Shahol et al.'s approach used a variable selection approach to select useful features from a CNN \cite{sahlol_efficient_2020}.  This approach selected over 1,000 variables to use for classifying the WBCs.  Our approach only required 24 features and outperformed their approach on both datasets.  This shows that our approach describes the WBCs in a more explainable and interpretable manner than the other approaches.  These results are also summarized in Tables \ref{tab: all_results} and \ref{tab: c_results}.



\subsection{Interpreting our models' results}

Table \ref{tab: all_VI_type} provides the relative variable importance (VI) by category.  The two most important categories are color and shape, while texture is the least important.  This differs from Bhattacharjee and Saini who used only shape features\cite{bhattacharjee_robust_2015}.  Sahlol \etal \ built a model using only texture features \cite{sahlol_automatic_2019}.  Singhal \etal \ built two different models: one that used only shape features and one that used only texture features.  Their final model suggested that the model using only the texture features was better\cite{singhal_local_2014}.  Rawat \etal \ built an ensemble of different models using color, shape, and texture features \cite{rawat_classification_2017}.  However, they provide no insight to what features were most important in their analysis.  Thus, our approach incorporates a smaller number of different types of features and outperforms all reported approaches.  Further, our models are the first characterize the importance of color for classifying WBCs as malignant or healthy.  Our results suggest that color is a major factor for classifying WBCs as malignant or healthy.

\begin{table}[h!]
\centering
\begin{tabular}{|c|cc|}
  \hline
    Relative Importance & ALL-IDB2 & C-NMC \\ 
  \hline
  Most              & Color: 1.000 & Shape: 1.000  \\ 
  Secondarily                 & Shape: 0.357 & Color: 0.776  \\ 
  Least                 & Texture: 0.052 & Texture: 0.152  \\ 
  
   \hline
\end{tabular}
\caption{The relative VI of the categories for the ALL-IDB2 and C-NMC data.}
\label{tab: all_VI_type}
\end{table}

Since we used interpretable features, we have greater insight into what is important for classifying malignant and healthy WBCs.  Both datasets indicate that color features are the most important, followed by shape and then texture.  However, if the WBC is extracted by an oncologist, shape features turn out to be very important for classifying the WBCs.  However, if the WBC must be extracted from the background, other features such as color tend to be more important.  Thus, the manner in which labs treat WBCs dramatically changes what features are useful for classification.  This in turn changes what the model learns.  

This also showcases that the RF model values different variables when trained on different datasets.  This is strong evidence of batch effects.  Deploying a RF model without considering batch effects is inadvisable.  Observing how the model performs alongside current systems is strongly encouraged before fully adopting any machine learning or artificial intelligence solution.

\section{Conclusion}

We provide a competitive and novel solution for WBC classification of Leukemia and healthy cells.  Our approach outperforms by about 4.38\% while also using substantially less features than other proposed solutions.  Further, our features are all explainable which provides greater insight to finding a solution to understand what characteristics are useful for classifying malignant and healthy WBCs.  Thus, we provide a pathway for a truly interpretable and explainable solution for classifying Leukemia and healthy WBCs.  Further, we were able to show that the variable importance differed for the different types of features.  This allowed us to show that if labs treat the WBCs differently, this has a direct impact on the classification models built.  Further, the importance of the variables used could easily change depending on the labs' treatments of their respective data.

\section*{Acknowledgment}

The Office of the Provost at George Mason University funded this research.

We would like to thank the Zang Lab for Computational Biology at the University of Virginia for their support. 

\bibliographystyle{IEEEtran}
\bibliography{zotero2.bib}

\clearpage 
\appendix
\section*{Segmentation}\label{app: seg}
The first step to extract the cell from our images.  This is crucially important and challenging since the ALL-IDB2 and C-NMC datasets were collected using different methods.  Furthermore, the ALL-IDB2 data includes the background surrounding the WBC.  Conversely, the C-NMC images have been segmented by an expert oncologist.  Examples of these images are found in Figure \ref{fig: overview}.   

Our image segmentation algorithm was partially inspired by Ghane et al. in the use of the CMYK color space, histogram equalization, and contrast stretching \cite{ghane_segmentation_2017}.  For a given image, $\textbf{a}[\vec{x}]$, we first extract the red, green, and blue channels, respectively, by \begin{equation}
    \textbf{r}[\vec{x}] = \left \{
    \begin{array}{c}
        1  \\
        \emptyset \\
        \emptyset
    \end{array}
\right \} \textbf{a}[\vec{x}] ,
\end{equation}

\begin{equation}
    \textbf{g}[\vec{x}] = \left \{
    \begin{array}{c}
        \emptyset  \\
        1 \\
        \emptyset
    \end{array}
\right \} \textbf{a}[\vec{x}],
\end{equation}

\begin{equation}
    \textbf{b}[\vec{x}] = \left \{
    \begin{array}{c}
        \emptyset  \\
        \emptyset \\
        1
    \end{array}
\right \} \textbf{a}[\vec{x}].
\end{equation} 

\noindent We need to then convert all of the black pixels to yellow.  This will allow us to extract our cell from the image more easily in the CMYK color space.  Let $v$ and $h$ be the width and height of image $\textbf{a}[\vec{x}]$.  Then $\forall i\in \{1, ..., v\}$ and $\forall j \in \{1, ..., h\}$, we convert the black pixels to yellow by \begin{equation}
    \textbf{r}^*_{i,j}[\vec{x}] = \left \{
    \begin{array}{ll}
        255, &  \textbf{r}_{i,j}[\vec{x}]\stackrel{?}{=}\textbf{g}_{i,j}[\vec{x}]\stackrel{?}{=}\textbf{b}_{i,j}[\vec{x}]\stackrel{?}{=} 0 \\
        \textbf{r}_{i,j}[\vec{x}], & \text{else}
    \end{array} \right.,
\end{equation}

\begin{equation}
    \textbf{g}^*_{i,j}[\vec{x}] = \left \{
    \begin{array}{ll}
        255, &  \textbf{r}_{i,j}[\vec{x}]\stackrel{?}{=}\textbf{g}_{i,j}[\vec{x}]\stackrel{?}{=}\textbf{b}_{i,j}[\vec{x}]\stackrel{?}{=} 0 \\
        \textbf{g}_{i,j}[\vec{x}], & \text{else}
    \end{array} \right.,
\end{equation}

\begin{equation}
    \textbf{b}^*_{i,j}[\vec{x}] = \left \{
    \begin{array}{ll}
        0, &  \textbf{r}_{i,j}[\vec{x}]\stackrel{?}{=}\textbf{g}_{i,j}[\vec{x}]\stackrel{?}{=}\textbf{b}_{i,j}[\vec{x}]\stackrel{?}{=} 0 \\
        \textbf{b}_{i,j}[\vec{x}], & \text{else}
    \end{array} \right.,
\end{equation} 

\begin{equation}
    \textbf{d}[\vec{x}] =   \left \{
    \begin{array}{c}
        \textbf{r}^*[\vec{x}] \\
        \textbf{g}^*[\vec{x}]\\
        \textbf{b}^*[\vec{x}]
    \end{array} \right \}.
\end{equation} 

\noindent Now that we converted all the black pixels to yellow, we can convert our input image to the CMYK color space and extract the Y channel by\begin{equation}
    \textbf{f}[\vec{x}] = \large{(} \left \{
    \begin{array}{c}
        \emptyset  \\
        \emptyset \\
        1 \\
        \emptyset
    \end{array}
\right \}\mathscr{L}_{CMYK}\textbf{d}[\vec{x}] \large{)} \otimes \textbf{g}[\vec{p}]
\end{equation}

\noindent where $\mathscr{L}_{CMYK}$ converts the input image to the CMYK color space, $\left \{
    \begin{array}{c}
        \emptyset  \\
        \emptyset \\
        1 \\
        \emptyset
    \end{array}
\right \}$ extracts the Y channel, $\otimes$ is the convolution operator, and $\textbf{g}[\vec{p}]$ is the local maximum filter using a $5\times5$ kernel. This process converts the image into a more useful color space, where we use the helpful channel for extracting our object of interest.  We also apply a local max operation to help remove some noise from the images.  

We then perform histogram equalization and contrast stretching to obtain two new images.  We obtain the contrast stretched image by \begin{equation}
    \textbf{s}[\vec{x}] = (\textbf{f}[\vec{x}] - \epsilon) \large{(}\frac{255}{\delta-\epsilon}\large{)},
\end{equation}
\noindent where $\delta$ and $\epsilon$ are the $2^{\text{nd}}$ and $98^{\text{th}}$ percentile values of image $\textbf{f}[\vec{x}]$. Next, to obtain the image with histogram equalization, we performed\begin{equation}
    \textbf{e}[\vec{x}] = \mathbb{E}\textbf{f}[\vec{p}],
\end{equation}

\noindent where $\mathbb{E}$ is the histogram equalization operator.  We then combined the two images as performed by Ghane et al. \cite{ghane_segmentation_2017}.  Explicitly, we combined the images by \begin{equation}
    \textbf{j}[\vec{x}] = (2\times\textbf{s}[\vec{x}]) + \textbf{e}[\vec{x}].
\end{equation}

\noindent We then performed a simple thresholding and extracted the object closest to the center of the image.  This was performed by \begin{equation}
    \textbf{u}[\vec{x}] = \mathcal{I}_{(v/2, h/2)}\Gamma_{(\bigwedge\textbf{j}[\vec{x}]+0.01)<}\textbf{j}[\vec{x}],
\end{equation}

\noindent where $\Gamma$ is the threshold operator and retains all the pixel values that are less than the the minimum value of image $\textbf{j}[\vec{x}]$ plus 0.01, $\bigwedge$ is the minimum value operator, $v$ and $h$ are the first and second values from $Z\textbf{j}[\vec{x}]$, respectively.  The size operator is $Z$.  The resulting image $\textbf{u}[\vec{x}]$ is a binary image where the object of interest is white (or 1's) and the background is black (or 0's).  This provides us with our extracted shapes as observed in Figure \ref{fig: overview}.
\section{Feature Extraction}\label{app: feat}

\subsection*{Shape}\label{app: shape}

We collected the encircled image-histograms (EIs) using the algorithm described by Lamberti \cite{lamberti_algorithms_2020}.  This algorithm results in a vector $\vec{c}_{EI}$ which contains the white and black pixel counts from image $ \textbf{u}[\vec{x}]$.  These counts are the first two metrics, $\vec{m}_1$ and $\vec{m}_2$, respectively.  The shape proportion (SP) value for a given image, $i$, is merely \begin{equation} \label{eq: other_fix}
    \vec{m}_{3, i} = \frac{\vec{m}_{1, i}}{\vec{m}_{1, i}+\vec{m}_{2, i}}.
\end{equation}

\begin{figure}[h!]
\centering
\includegraphics[scale=0.25]{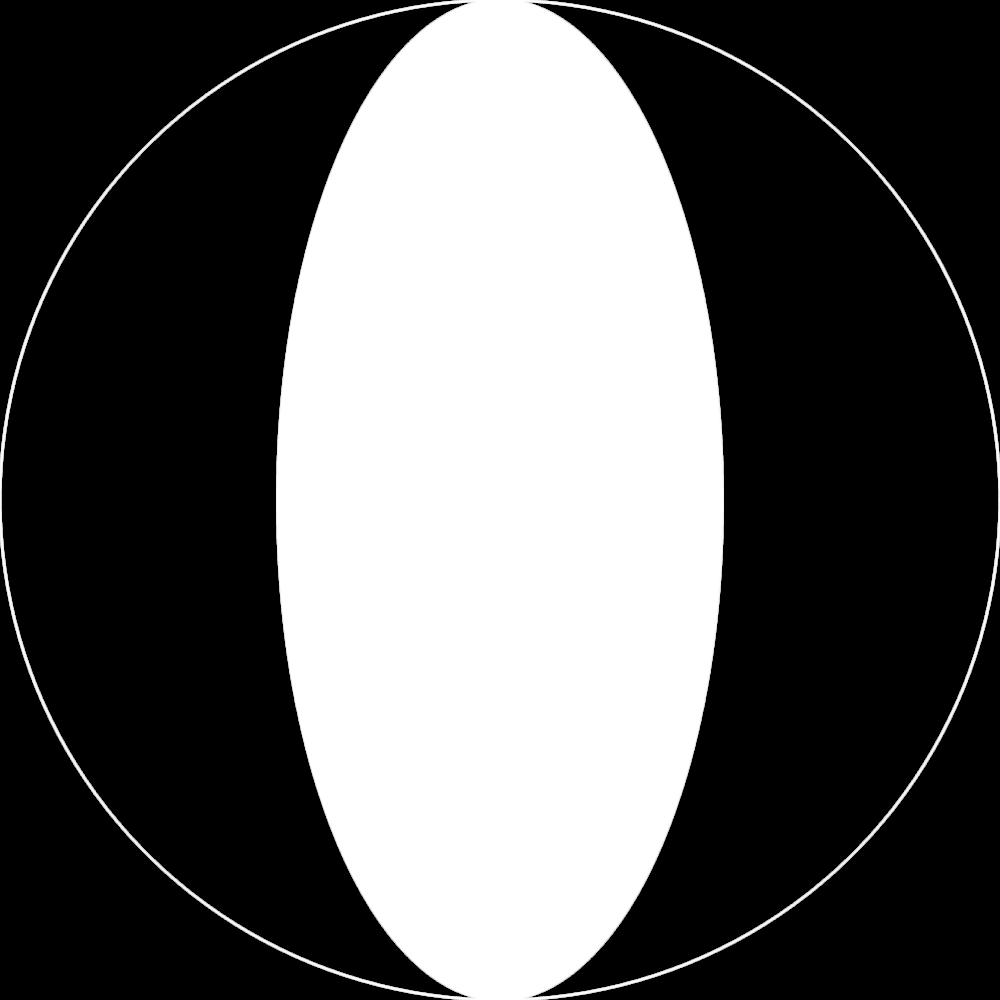}
\caption{ Figure provides an example of the SPEI algorithm result with a reference circle.  The SPEI algorithm will put the oval in the minimum encompassing circle.  Then SPEI will put the region in the minimum encompassing square.  The reference circle is only for presentation purposes.  It is not part of the actual final image. }
\label{fig: spei_ex}
\end{figure}

\noindent \noindent The SP value is essentially the proportion of white pixels after applying the SPEI algorithm \cite{lamberti_algorithms_2020}.  This SPEI algorithm puts a shape in its minimum encompassing circle.  Then the circle is placed in its minimum encompassing square.  Figure \ref{fig: spei_ex} provides an illustrative example of the SPEI algorithm result.  The EIs are the white and black pixel counts after applying SPEIs \cite{lamberti_algorithms_2020}.

We calculated eccentricity by finding the ratio of the first and second eigenvalues.  To obtain the eigenvalues, we performed \begin{equation} \label{eq: eigens}
    \vec{e}_{i} = \mathbb{E}_{(1,2)}V\textbf{u}_i[\vec{x}],
\end{equation}

\noindent where $V$ collects the covariance matrix of the shape matrix and $\mathbb{E}_{(1,2)}$ calculates the first and second eigenvalues of the resulting covariance matrix. The $j^{\text{th}}$ eigenvalue on image $i$ is $\vec{e}_{i,(j)}$.  To obtain $\vec{m}_6$, eccentricity, we perform on a given image, $i$,\begin{equation}
\label{eq: ecc}
     \vec{m}_{6, i} = \frac{\vec{e}_{i, (1)}}{\vec{e}_{i,(2)}}.
\end{equation} 

\noindent \noindent The first and second eigenvalues are metrics $ \vec{m}_{4}$ and $ \vec{m}_{5}$, respectively.  It is well-known that the eigenvalues of a covariance matrix correspond to the linear combination in the data which maximizes the variance for their respective dimension \cite{izenman_modern_2008}.  For instance, the first eigenvalue is the linear combination of the data which maximizes the first eigenvalue \cite{izenman_modern_2008}.  We also know that the linear projections, or eigenvalues, are orthogonal to one another \cite{izenman_modern_2008}.  Thus, the eigenvalues are measures of the major and minor axes of our given shape.  Using the ratio of the major and minor axes provides some insight to how a given shape exists as a 2D digital image \cite{kinser_image_2018}.  A value of 1 corresponds to a shape with the same major and minor axes' lengths.  A value greater than 1 corresponds to the case where the major axis' length is larger than the minor axis' length.  The limit of eccentricity would correspond to the case where the major axis' length is infinitely larger than the minor axis' length.  

Circularity, $\vec{m}_{7}$, was collected on image $i$ by
\begin{equation}
\label{eq: circ}
     \vec{m}_{7,i} = \frac{\sum \textbf{u}_i[\vec{x}]}{4\pi\times\big( \sum\triangleleft\textbf{u}_i[\vec{x}]-\sum \textbf{u}_i[\vec{x}] \big)}, 
\end{equation} 
\noindent \noindent where $\sum$ sums the pixel intensity values.  The $\sum$ operator will compute the area of the image since we are restricted to binary images.  The denominator of this metric is the perimeter of the binary image multiplied by $4\pi$.  Circularity provides a measure for how circular a shape is as a 2D digital image \cite{kinser_image_2018, rosenfeld_compact_1974, proffitt_measurement_1982}.  A value of 1 corresponds to a perfect circle \cite{kinser_image_2018, rosenfeld_compact_1974, proffitt_measurement_1982}.  

The number of corners, $\vec{m}_{8}$, counts the number of corners on an object \cite{harris_combined_1988}.  This metric was measures on image $i$ by 
\begin{equation}
    \textbf{v}_i[\vec{x}] = G_2E_{x}\textbf{u}_i[\vec{x}],
\end{equation}
\begin{equation}
    \textbf{h}_i[\vec{x}] = {G}_2{E}_{y}\textbf{u}_i[\vec{x}],
\end{equation}
\begin{equation}
    \textbf{A}_i[[\vec{x}]] = \begin{bmatrix}
         \textbf{v}_i[\vec{x}]^2 & \textbf{v}_i[\vec{x}]\textbf{h}_i[\vec{x}]\\
         \textbf{v}_i[\vec{x}]\textbf{h}_i[\vec{x}] & \textbf{h}_i[\vec{x}]^2
         \end{bmatrix},
\end{equation}

\begin{equation}
    \textbf{r}_i[\vec{x}] = \frac{det(\textbf{A}_i[[\vec{x}]]) - k}{ trace(\textbf{A}_i[[\vec{x}]])^2 },
\end{equation}

\begin{equation}
    \vec{m}_{8,i} = \mathbb{C} \textbf{r}_i[\vec{x}] ,
\end{equation}

\noindent \noindent where $E$ is the edge operator where the subscript finds the vertical or horizontal edges, $x$ and $y$, respectively, $G_2$ is a Gaussian smoothing operator with a standard deviation of 2, $\textbf{A}_i[[\vec{x}]]$ is an image tensor, $\textbf{r}_i[\vec{x}]$ are the resulting corners extracted from the image, and $\mathbb{C}$ extracts the number of corners in the input image.  Here we first extract the edges from the image to help first isolate the parts of the image that would contain the corners.  Next, we smooth the image to remove spurious parts of the image that would not qualify as a corner.  We then create a tensor and compute image $\textbf{r}_i[\vec{x}]$ to create an image that contains all of the corners in the image.  We then count the number of corners as the last operator to retrieve the desired metric.

Note that all of the variables used in our analysis have a very interpretable meaning.  The EI's are the white and black pixel counts after putting the shape in the minimal encompassing circle and then the minimal encompassing square.  The SP value is the proportion of white pixels divided by the summation of the EIs.  The $1^{\text{st}}$ and $2^{\text{nd}}$ Eigenvalues are associated with the major and minor axes, respectively, of the shape.  Eccentricity is the ratio of the major to minor axis.   Circularity measures how circular a shape is where a value of 1 indicates a perfect circle.  The number of corners estimates the number of corners the shape has.  

\subsection*{Color}\label{app: color}

The features for color were the mean and standard deviation (SD) of the object's pixel intensity values.  The object's pixels were extracted by using the segmented shape.  Seven different color channels were used from two different color representations: RGB and CMYK.  Explicitly, for a given color channel and image, metrics $\vec{m}_{9,i}$ to $\vec{m}_{22,i}$\begin{equation}
    \vec{m}_{9,i} = \mathcal{M}_{\textbf{u}_i[\vec{x}]} \textbf{r}_i[\vec{x}],
\end{equation}
\begin{equation}
    \vec{m}_{10,i} = \mathcal{T}_{\textbf{u}_i[\vec{x}]} \textbf{r}_i[\vec{x}],
\end{equation}
\begin{equation}
    \vec{m}_{11,i} = \mathcal{M}_{\textbf{u}_i[\vec{x}]} \textbf{g}_i[\vec{x}],
\end{equation}
\begin{equation}
    \vec{m}_{12,i} = \mathcal{T}_{\textbf{u}_i[\vec{x}]} \textbf{g}_i[\vec{x}],
\end{equation}
\begin{equation}
    \vec{m}_{13,i} = \mathcal{M}_{\textbf{u}_i[\vec{x}]} \textbf{b}_i[\vec{x}],
\end{equation}
\begin{equation}
    \vec{m}_{14,i} = \mathcal{T}_{\textbf{u}_i[\vec{x}]} \textbf{b}_i[\vec{x}],
\end{equation}
\begin{equation}
    \vec{m}_{15,i} = \mathcal{M}_{\textbf{u}_i[\vec{x}]}  \large{(} \left \{
    \begin{array}{c}
        1  \\
        \emptyset \\
        \emptyset \\
        \emptyset
    \end{array}
\right \}\mathscr{L}_{CMYK}\textbf{d}[\vec{x}] \large{)},
\end{equation}
\begin{equation}
    \vec{m}_{16,i} = \mathcal{T}_{\textbf{u}_i[\vec{x}]}  \large{(} \left \{
    \begin{array}{c}
        1  \\
        \emptyset \\
        \emptyset \\
        \emptyset
    \end{array}
\right \}\mathscr{L}_{CMYK}\textbf{d}[\vec{x}] \large{)},
\end{equation}
\begin{equation}
    \vec{m}_{17,i} = \mathcal{M}_{\textbf{u}_i[\vec{x}]}  \large{(} \left \{
    \begin{array}{c}
        \emptyset  \\
        1 \\
        \emptyset \\
        \emptyset
    \end{array}
\right \}\mathscr{L}_{CMYK}\textbf{d}[\vec{x}] \large{)},
\end{equation}
\begin{equation}
    \vec{m}_{18,i} = \mathcal{T}_{\textbf{u}_i[\vec{x}]}  \large{(} \left \{
    \begin{array}{c}
        \emptyset  \\
        1 \\
        \emptyset \\
        \emptyset
    \end{array}
\right \}\mathscr{L}_{CMYK}\textbf{d}[\vec{x}] \large{)},
\end{equation}
\begin{equation}
    \vec{m}_{19,i} = \mathcal{M}_{\textbf{u}_i[\vec{x}]}  \large{(} \left \{
    \begin{array}{c}
        \emptyset  \\
        \emptyset \\
        1 \\
        \emptyset
    \end{array}
\right \}\mathscr{L}_{CMYK}\textbf{d}[\vec{x}] \large{)},
\end{equation}
\begin{equation}
    \vec{m}_{20,i} = \mathcal{T}_{\textbf{u}_i[\vec{x}]}  \large{(} \left \{
    \begin{array}{c}
        \emptyset  \\
        \emptyset \\
        1 \\
        \emptyset
    \end{array}
\right \}\mathscr{L}_{CMYK}\textbf{d}[\vec{x}] \large{)},
\end{equation}
\begin{equation}
    \vec{m}_{21,i} = \mathcal{M}_{\textbf{u}_i[\vec{x}]}  \large{(} \left \{
    \begin{array}{c}
        \emptyset  \\
        \emptyset \\
        \emptyset \\
        1
    \end{array}
\right \}\mathscr{L}_{CMYK}\textbf{d}[\vec{x}] \large{)},
\end{equation}
\begin{equation}
    \vec{m}_{22,i} = \mathcal{T}_{\textbf{u}_i[\vec{x}]}  \large{(} \left \{
    \begin{array}{c}
        \emptyset  \\
        \emptyset \\
        \emptyset \\
        1
    \end{array}
\right \}\mathscr{L}_{CMYK}\textbf{d}[\vec{x}] \large{)},
\end{equation}

\noindent where $\mathcal{M}$ and $ \mathcal{T}$ are the mean and standard deviation operators, respectively.  Both operators are only extracted on the 1's of image $\textbf{u}[\vec{x}]$, which ensures that only the colors of object are considered for a given color channel.  The mean value for a given channel indicates how much of that color is in the respective channel.  Large values indicate that a particular color is very present in the image, while small correspond to the image having very little of a given color.  The SD indicates how much the presence of a color changes throughout an image.  A large SD indicates that the presence of the intensity of a color changes dramatically, while small a SD corresponds to a fairly uniform representation of the intensity of a color.  Thus, our color features have an interpretable and explainable meaning.

\subsection*{Texture}\label{app: text}
The features for texture where the mean and SD of the co-occurance matrix of the masked grayscale image.  This was claculated by:
\begin{equation}
    \vec{m}_{23,i} = \mathcal{M} \mathfrak{C}_{1, 256} {\large{(}}(\mathscr{L}_L\textbf{a}_i[\vec{x}]) \times \textbf{u}_i[\vec{x}]{\large{)}},
\end{equation}
\begin{equation}
    \vec{m}_{24,i} = \mathcal{T} \mathfrak{C}_{1, 256} {\large{(}}(\mathscr{L}_L\textbf{a}_i[\vec{x}]) \times \textbf{u}_i[\vec{x}]{\large{)}},
\end{equation}

\noindent \noindent where $\mathscr{L}_L$ converts the given input image to a grayscale image and $\mathfrak{C}$ calculates the co-occurrence matrix with a shift of 1 to the right and bottom for 256 intensities.  We multiply the grayscaled image by the extracted shape to remove the background from the image.  We then calculate the co-occurrence matrix.  This image provides the counts of number of times a certain pixel intensity is next to another intensity with a single shift.  This $256\times256$ matrix is useful for texture measures \cite{kinser_image_2018}.  We then extracted the mean and SD from this matrix.  The mean of this matrix indicates how often an average intensity is nearby another value.  Large means would indicate that intensities are nearby other intensity values often which translates to a very rough image.  Small means would indicate that intensities are nearby very few other values with translates to a smoother image.  The SD indicates how often much the intensities change over an image, with large values indicating a large amount of change and small values indicating a small amount of change.  Thus, our texture features have an interpretable and explainable meaning.

\section*{Extended Results}

Confusion matrices of the ALL-IDB2 training and validation data are provided in Tables \ref{tab: all_conf_train} and \ref{tab: all_conf_valid}, respectively.  Confusion matricies of the C-NMC training and validation data are provided in Tables \ref{tab: cnmc_conf_train} and \ref{tab: cnmc_conf_valid}.  In Tables \ref{tab: all_conf_train}, \ref{tab: all_conf_valid}, and \ref{tab: cnmc_conf_train}, we were able to obtain perfect classification.  Table \ref{tab: cnmc_conf_valid} shows that our RF model for the C-NMC data did overfit, but was still able to perform well as the Tables have large values in the diagonal.  Future work would include producing a model that did not overfit as much to the training data but also improved on the validation data.  

Table \ref{tab: rel_VI} provides the two relative VI for the two datasets.  Table \ref{tab: VI} compares the VI for the two datasets.  Tables \ref{tab: rel_VI} and \ref{tab: VI} show that the VI changes between the two datasets.  This provides strong evidence in the batch effects between these two datasets.  Thus, analysts must consider batch effects when using models in deployment. 

\begin{table}[h!]
\centering
\begin{tabular}{|c|c|c|}
\hline
Prediction/Truth & Healthy (Negative) & Malignant (Positive) \\
\hline
 Healthy (Negative)  & 104& 0   \\ \hline
 Malignant (Positive)& 0 & 104   \\ \hline
\end{tabular}
\caption{The ALL-IDB2 confusion matrix for the training data.}
\label{tab: all_conf_train}
\end{table}

\begin{table}[ht]
\centering
\begin{tabular}{|c|c|c|}
\hline
Prediction/Truth & Healthy (Negative) & Malignant (Positive) \\
\hline
 Healthy (Negative)  & 26& 0   \\ \hline
 Malignant (Positive)& 0 & 26   \\ \hline
\end{tabular}
\caption{The ALL-IDB2 confusion matrix for the validation data.}
\label{tab: all_conf_valid}
\end{table}

\begin{table}[h!]
\centering
\begin{tabular}{|c|c|c|}
\hline
Prediction/Truth & Healthy (Negative) & Malignant (Positive) \\
\hline
 Healthy (Negative)  & 2711& 0   \\ \hline
 Malignant (Positive)& 0 & 5817   \\ \hline
\end{tabular}
\caption{The C-NMC confusion matrix for the training data.}
\label{tab: cnmc_conf_train}
\end{table}

\begin{table}[h!]
\centering
\begin{tabular}{|c|c|c|}
\hline
Prediction/Truth & Healthy (Negative) & Malignant (Positive) \\
\hline
 Healthy (Negative)  & 522 & 133    \\ \hline
 Malignant (Positive)& 156 & 1322   \\ \hline
\end{tabular}
\caption{The C-NMC confusion matrix for the validation data.}
\label{tab: cnmc_conf_valid}
\end{table}

\begin{table*}[h!]
\centering
\begin{tabular}{cc||cc}
  \hline
 Feature & Relative VI & Feature & Relative VI  \\ \hline
  Mean M & 1.000 &    White EI & 1.000 \\ 
  SD K & 0.791 &  $2^{\text{nd}}$Eigenvalue & 0.664 \\  
  SD B & 0.660 &  $1^{\text{st}}$Eigenvalue & 0.614 \\
  $2^{\text{nd}}$Eigenvalue & 0.539 &  Mean B & 0.436\\ \\
  White EI & 0.458 &  Black EI & 0.427 \\
  Mean G & 0.393 &  Mean K & 0.414 \\
  SD M & 0.253 &  Mean R & 0.265 \\
  Mean K & 0.229 &  Mean P & 0.250 \\ \\
  Mean B & 0.214 &  Mean G & 0.248 \\
  SD G & 0.209 & SD M & 0.247 \\ 
  Circularity & 0.207 &  SD P & 0.245 \\
  SD C & 195 &  Number of Corners & 0.204 \\ \\
  $1^{\text{st}}$Eigenvalue & 0.146 &  SD K & 0.176 \\
  SD R & 0.139 &  Mean M & 0.175 \\
  Mean P & 0.124 &  SD B & 0.165 \\
  Number of Corners & 0.105 &  Circularity & 0.162 \\ \\
  SD P & 0.097 &  SD G & 0.154 \\
  Mean R & 0.087 &  SP & 0.129 \\
  SD C & 0.066 &  Mean C & 0.103 \\
  SP & 0.057 &  SD R & 0.088 \\ \\
  Black EI & 0.052 &  Eccentricity & 0.060 \\
  Eccentricity & 0.023 &  SD C & 0.056\\
  Mean Y & 0.001 &  Mean Y & $>$0.000 \\
  SD Y & $>$0.000 &  SD Y & $>$0.000 \\
   \hline
\end{tabular}
\caption{The table provides the relative VI of the features for the ALL-IDB2 data in the first two columns and the C-NMC data in the second two columns.  The most important feature for the ALL-IDB2 was the mean pixel value in the blue channel, followed closely by the mean pixel value in the black channel and then circularity.  The most important feature for the C-NMC data was the White EI value, followed closely by the Black EI and Eccentricity.  Each of these variables is defined in the Interpretable Feautres subsection of the Methods section.}
\label{tab: rel_VI}
\end{table*}

\begin{table*}[ht!]
\centering
\begin{tabular}{c|cc}
  \hline
Feature & Mean Decrease in Accuracy for ALL-IDB & Mean Decrease in Accuracy for C-NMC \\ \hline
  \hline
  White EI & 0.039       & 0.083 \\ 
  Black EI & 0.004       & 0.035 \\ 
  SP       & 0.005       & 0.011 \\ \\
  Mean R   & 0.007       & 0.022 \\ 
  SD R     & 0.012      & 0.007 \\ 
  Mean G   & 0.033       & 0.021 \\ 
  SD G     & 0.018       & 0.013\\ 
  Mean B   & 0.018      &  0.036 \\ 
  SD B     & 0.056       &  0.014\\ \\
  Mean C   & 0.017       & 0.009\\ 
  SD C     & 0.006       &  0.005\\ 
  Mean M   & 0.085       &  0.015\\ 
  SD M     & 0.021       &  0.021\\ 
  Mean Y   & $>$0.000       &  $>$0.000\\ 
  SD Y     & $>$0.000       &  $>$0.000\\ 
  Mean K   & 0.019       &  0.034\\ 
  SD K     & 0.067       &  0.015\\ \\
  Mean P   & 0.010       &  0.021\\ 
  SD P     & 0.008       &  0.020\\ \\
  Circularity                & 0.018        & 0.013 \\ 
  Eccentricity               & 0.002        & 0.005 \\ 
   $1^{\text{st}}$Eigenvalue & 0.012        & 0.051 \\ 
   $2^{\text{nd}}$Eigenvalue & 0.046        & 0.055 \\ 
  Number of Corners          & 0.009        & 0.017 \\ 
   \hline
\end{tabular}
\caption{The VI of the features for the ALL-IDB2 and C-NMC data.}
\label{tab: VI}
\end{table*}

\end{document}